\documentstyle[times,eqsecnum,aps,prd]{revtex}
\input epsf
\begin{document}


\wideabs{ 
\title{Analytic Quantization of the QCD String} 
\author{Theodore J. Allen}
\address{Physics Department, Hobart \& William Smith Colleges \\
Geneva, New York 14456 USA}

\author{Charles Goebel {\it and\/} M. G. Olsson}
\address{Department of Physics, University of Wisconsin, \\
1150 University Avenue, Madison, Wisconsin 53706 USA }

\author{Sini\v{s}a Veseli}
\address{Fermi National Accelerator Laboratory \\
P.O. Box 500, Batavia, Illinois 60510 USA}

\date{\today}
\maketitle

\begin{abstract}
We perform an analytic semi-classical quantization of the straight QCD
string with one end fixed and a massless quark on the other, in the limits
of orbital and radial dominant motion.  We compare our results to the exact
numerical semi-classical quantization.  We observe that the numerical
semi-classical quantization agrees well with our exact numerical canonical
quantization.
\end{abstract}
}  

\section{Introduction}\label{sec:intro}

The purpose of this paper is to explore some remarkable results of the QCD
string/flux tube model\cite{ref:one,ref:two,ref:three}.  In its
relativistic and single quantized form \cite{ref:three} a particularly
simple pattern emerges when one or both quarks are light.  For the purposes
of this paper we take the light quarks to have zero mass.  In this case the
energy $E$ of the light degrees of freedom (LDF) and the angular and radial
quantum numbers ($J$ and $n$ respectively) are accurately related by
\begin{equation}\label{eq:one}
{E^2 \over (2) \pi a} \simeq J + 2n + \frac32 
\end{equation}
for one or (two) light quark(s).  For the case of one light quark the
meson energy is the sum of $E$ and the heavy quark mass.

The above pattern of angular and radial states results in degenerate
``towers'' of mesons of the same parity.  This is the same pattern as the
3D harmonic oscillator.

The QCD string model is kinematically intricate and it would seem unlikely
that it would lead to such a simple result as Eq.~(\ref{eq:one}).  We
demonstrate in this paper that although Eq.~(\ref{eq:one}) is not exact, it
is very accurate for most accessible quantum states.  The quantized
relativistic flux tube model is of great interest because of the
probability that QCD reduces to string-like behavior at large source
separations \cite{ref:four}.

The simplest version of a quark string model assumes that the string is
always straight.  In the limiting cases of circular motion or pure radial
motion this assumption is physically reasonable.  In addition, for massive
quarks at the ends the relativistic corrections have been shown
\cite{ref:five} to agree with the Wilson loop description of QCD
confinement \cite{ref:six}.  Numerical quantization of the straight string
and quark system has been done canonically \cite{ref:three} with the
Nambu-Goto string, as well as in the WKB approximation with an auxiliary
field method \cite{ref:seven}.  Both approaches give results similar to
Eq.~(\ref{eq:one}).

More generally, one may allow the string to curve adiabatically by
incorporating the string equations of motion from the Nambu-Goto action
\cite{ref:eight}.  This more general string calculation demonstrates that
the string curvature remains small for motion in ordinary hadrons.  The
string curvature only slightly changes the energies of the bound states,
justifying the use of the simpler straight string approximation.

The approach emphasized here is to quantize the straight string system
semi-classically.  We will show that this quantization agrees well with our
previous exact quantization method.  We will also show that a single
integral function then predicts the whole spectroscopy when at least one
quark is massless.  We will approximate analytically this integral in
sectors when $J \gg n$ and when $n \gg J$.  The later case where radial
motion dominates is valid over most of the allowed bound states.

\section{Dynamics and Quantization}\label{sec:dynamics}

\subsection{Dynamics}\label{subsec:dynamics}
As we mentioned earlier, the straight QCD string is an excellent
approximation to the dynamical (curved) string in normal hadrons.  We will
therefore restrict ourselves to the relatively simple straight string.  We
will explicitly consider the case of one fixed end and a quark of mass $m$
at the other. The string with two light quarks introduces only minor
modifications, which we discuss at the end of section \ref{sec:dynamics}.
As is well-known \cite{ref:three}, the two constants of motion are the
orbital angular momentum $J$ and the energy $E$ of the light degrees of
freedom, which are given in terms of the quark's transverse velocity
$v_\perp$ and radial momentum $p_r$ as
\begin{eqnarray}
J & = & W_r \gamma_\perp v_\perp r + a r^2 f(v_\perp) \ , \label{eq:Jdef} \\
E & = & W_r \gamma_\perp + a r S(v_\perp) \ , \label{eq:Mdef} 
\end{eqnarray}
where $W_r = \sqrt{p_r^2 + m^2}$, $\gamma_\perp = (1 - v_\perp^2)^{-1/2}$,
$r$ is the string length, and $a$ is the string tension.  The functions $f$
and $S$ that appear in the expressions for the angular momentum and energy
are
\begin{eqnarray}
f(v_\perp) & = & {1\over 2 v_\perp}\left(S(v_\perp) - \sqrt{1 -
v_\perp^2}\right) \ , \label{eq:fdef} \\ S(v_\perp) & = &
{\sin^{-1}(v_\perp) \over v_\perp} \ . \label{eq:Sdef}
\end{eqnarray}

For our present purposes we will introduce a set of dimensionless
variables.  As our units we take the circular orbit radius (in the limit of
a massless quark)
\begin{equation}
r_0 = 2\sqrt{J\over \pi a} \ , \label{eq:r0}
\end{equation}
and corresponding string energy 
\begin{equation}
E_0 = \sqrt{J\pi a} \ .  \label{eq:M0}
\end{equation}

Our dimensionless variables $\Delta$, $\rho$, and $w$ are defined  by
\begin{eqnarray}
{E\over E_0} & \equiv & \sqrt{1 + \Delta^2} \ , \label{eq:M} \\
{r\over r_0} & \equiv & \sqrt{1 + \Delta^2} + \rho\Delta \ , \label{eq:r}
\end{eqnarray}
and
\begin{eqnarray}
w & \equiv & { W_r \over E_0 } \ .
\end{eqnarray}
The leading (classical) Regge trajectory corresponds to $\Delta = 0$ and
radial excitation occurs for positive $\Delta$.

In terms of these dimensionless variables the conserved quantities
(\ref{eq:Jdef}) and (\ref{eq:Mdef}) become
\begin{eqnarray}
1 & = & 2\, w \gamma_\perp \, v_\perp \left[\sqrt{1 + \Delta^2} + \rho\Delta\right]
\nonumber \\ 
& & + {4\over \pi} f(v_\perp) \left[\sqrt{1 + \Delta^2} + \rho\Delta\right]^2 \ , \\
\sqrt{1 + \Delta^2} & = & w\gamma_\perp + {2\over \pi} S(v_\perp)
\left[\sqrt{1 + \Delta^2} + \rho\Delta\right] \ \label{eq:Mtoo}.
\end{eqnarray}

After some work, we can eliminate the product $w\gamma_\perp$ to obtain
\begin{eqnarray}
& 2& v_\perp\left[\sqrt{1 + \Delta^2} + \rho\Delta\right]\sqrt{1 + \Delta^2}
 \nonumber 
\\ & - & \, {4\over \pi}\left(v_\perp S(v_\perp) - f(v_\perp)\right)
\left[\sqrt{1 + \Delta^2} + \rho\Delta\right]^2 
\nonumber \\
& =& 1 \label{eq:wgam}
\end{eqnarray}
and rewrite Eq.~(\ref{eq:Mtoo}) to find an expression for $w$
\begin{eqnarray}
w & = & \sqrt{1-v^2_\perp}\Bigg[\sqrt{1 + \Delta^2} \nonumber \\ 
 & & - {2\over \pi}
S(v_\perp) \left(\sqrt{1 + \Delta^2} + \rho\Delta\right)\Bigg] \ . \label{eq:w}
\end{eqnarray}

At the radial turning points $p_r = 0$.  The radial velocity also vanishes,
except in the massless quark limit.

The reason for the definition of the radial coordinate $\rho$ in
Eq.~(\ref{eq:r}) will be evident in this limit.  If we take the limit
$v_\perp = 1$ this will correspond to $\dot r = 0$ in the massless quark
case.  Using $f(1) = \pi/4$ and $S(1) = \pi/2$ in Eq.~(\ref{eq:wgam}) we
see that the solution is $\rho^2 = 1$ for any $\Delta$.  The points $\rho =
\pm 1$ therefore correspond to possible turning points.

\begin{figure}[hbp]
\epsfxsize = \linewidth 
\hspace*{-5mm}
\epsfbox{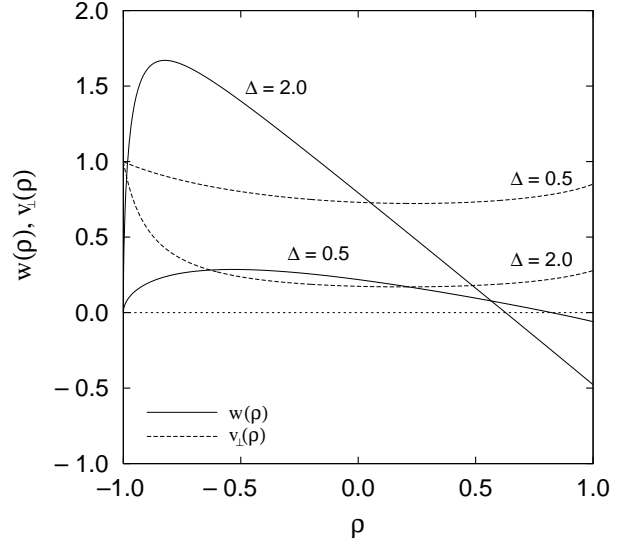}
\vskip 1 cm
\caption{Exact numerical solutions of Eq.~(\protect\ref{eq:wgam}) and
Eq.~(\protect\ref{eq:w}) for $w(\rho)$ and $v_\perp(\rho)$ for the values
$\Delta =0.5$ and $\Delta = 2.0$.}
\label{fig:one}
\end{figure}

By Eq.~(\ref{eq:w}) we see that at $v_\perp = 1$, $w=0$.  This would seem
to be the expected result in the massless quark case.  However, as we see
in Fig.~\ref{fig:one}, the second factor in Eq.~(\ref{eq:w}) has a zero at
$\rho < 1$ and is the true turning point.  In Fig.~\ref{fig:one} we show
representative numerical solutions with $\Delta = 0.5$ and $\Delta=2.0$.
First we solve Eq.~(\ref{eq:wgam}) for $v_\perp$ for $-1<\rho< +1$ and
substitute into Eq.~(\ref{eq:w}) to compute $w(\rho)$.  The turning points
of the general motion are $\rho_- = -1$ and $\rho_+ < 1$.  We will show
later that in the massless quark case
\begin{equation}
{\pi \over 2} -1 \leq \rho_+ \leq 1 \ , 
\end{equation}
where the lower limit is achieved for large $\Delta$ and $\rho_+ = 1$ when
$\Delta = 0$.

The outer turning point in the massless case is a ``bounce'' where
$\dot{r}$ discontinuously changes sign.

\subsection{Quantization} \label{subsec:quantization} 

The semi-classical quantization condition for a spherically symmetric
system is \cite{ref:nine}
\begin{equation}
\int_{r_-}^{r_+}\kern-5pt dr\,\, p_r\left(r,J+1/2\right) = \pi\left(n +
\frac12\right) \ ,
\end{equation}
where $n=0,1,2,\ldots$ and it is understood that the Langer correction
\cite{ref:nine} replaces the classical angular momentum $J$ by $J+ 1/2$,
where $J$ is now the angular momentum quantum number.  In terms of our
dimensionless parameters $\rho$ and $w$, the above quantization condition
becomes
\begin{equation}
{4 \over \pi }\,\int_{-1}^{\rho_+} d\rho\, \sqrt{w^2 - {m^2 \over (J +
\frac12)a\pi}}  = {(2n+1)\over (J + \frac12)\Delta} \ . \label{eq:semic}
\end{equation}

For our purposes, the massless quark will be of central interest.  In this
case, there is no $J$ dependence on the left hand side of
Eq.~(\ref{eq:semic}) and, by defining
\begin{equation}\label{eq:I}
I(\Delta) = {4\over \pi} \int_{-1}^{\rho_+} \kern-5pt d\rho\,\, w(\rho) \ ,
\end{equation}
we obtain
\begin{equation}\label{eq:Qcond}
\Delta\, I(\Delta) = {2n + 1 \over J + \frac12} \ . 
\end{equation}

\begin{figure}[hbp]
\epsfxsize = \linewidth 
\hspace*{-5mm}
\epsfbox{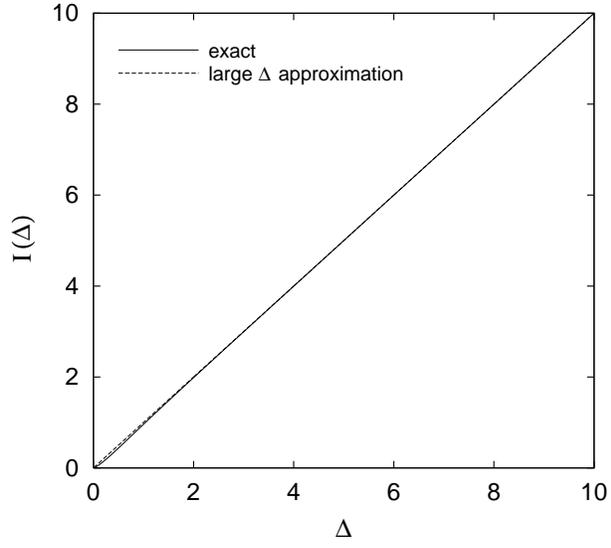}
\vskip 1 cm
\caption{A comparison of the exact numerical solution of
Eq.~(\protect\ref{eq:I}) for $I(\Delta)$ with the leading order approximation
$I(\Delta)\simeq \Delta$. }
\label{fig:two}
\end{figure}

The remarkable aspect of the massless quantization condition is that the
whole spectrum is revealed once $I(\Delta)$ is computed.  In
Fig.~\ref{fig:two}, we show the numerical result of this integration.  For
a given $\Delta$ and $\rho$ we use Eq.~(\ref{eq:wgam}) to find $v_\perp$.
The result is used in Eq.~(\ref{eq:w}) to find $w(\rho)$.  This is turn is
used in Eq.~(\ref{eq:I}) to compute $I(\Delta)$.  The upper integration
limit is determined by the zero in $w(\rho)$, as shown in
Fig.~\ref{fig:one}.

It is evident from Fig.~\ref{fig:two} that $I(\Delta)$ is essentially
linear in $\Delta$ with unit slope.  The LDF energy dependence is related
to $\Delta$ by Eq.~(\ref{eq:M})
\begin{equation}\label{eq:Deltatoo}
\Delta = \Bigg[{E^2 \over (J + \frac12) a \pi} - 1\Bigg]^{1/2} \ , 
\end{equation}
where we have again used the Langer correction to the angular momentum.
Using the quantization condition, Eq.~(\ref{eq:Qcond}), we can map out the
entire Regge structure as shown by the curves in Fig.~\ref{fig:three}.  The
trajectories are labeled by different radial excitations
$n=0,1,2,\ldots$. The $n=0$ trajectory is normally called the ``leading
trajectory'' and the $n>0$ trajectories are known as the ``daughter
trajectories.''  The solid points are the exact numerical solutions by the
exact canonical quantization method \cite{ref:three}.  The nearly perfect
agreement between the semi-classical and canonical quantization should be
noted.  It is only along the leading trajectory that small differences
arise.  If there are differences, this is where they should appear since the
semi-classical quantization becomes exact for large radial excitation where
the wave function has many nodes.

Figure \ref{fig:two} shows that for $\Delta >1$,  $I(\Delta)$ is essentially
equal to $\Delta$.  In general we expect an expansion in $\Delta^{-1}$ to
have odd powers of the form
\begin{equation} \label{eq:Deltaexp}
I(\Delta) \buildrel \Delta \gg 1 \over \longrightarrow \Delta + {c \over
\Delta} + \ldots \ .
\end{equation}
A detailed examination of our exact numerical integration of
Eq.~(\ref{eq:I}) shows that 
\begin{equation}\label{eq:numc}
c \simeq -0.006 \ .
\end{equation}
The small magnitude of the coefficient $c$ reflects the extraordinary  
accuracy of $I(\Delta)=\Delta$ for large Delta.

The string with two light quarks can be considered as two single light
quark strings with their fixed ends coinciding at the center of mass point.
The resulting light meson equations following from Eqs.~(\ref{eq:Jdef}) and
(\ref{eq:Mdef}) are 
\begin{eqnarray}
J_L & = & W_r \gamma_\perp v_\perp r_L + \frac12 a r_L^2 f(v_\perp) \ , \label{eq:J2light} \\
E_L & = & 2W_r \gamma_\perp + a r_L S(v_\perp) \ , \label{eq:M2light} 
\end{eqnarray}
where $r_L = 2r$, $J_L = 2 J$, and $E_L = 2E$. 

We recover Eq.~(\ref{eq:wgam}) upon rescaling
\begin{eqnarray}
E_{0L} & = & \sqrt{2\pi a J_L} \ , \\
J_{0L} & = & 2 \sqrt{2 J_L \over a \pi} \ .
\end{eqnarray}

Using Eq.~(\ref{eq:w}), we may now identify 
\begin{equation}
W_r \equiv \frac12 \sqrt{2\pi a J_L} \,\, w(\rho)\ , 
\end{equation}
and the quantization condition with $I(\Delta) \simeq \Delta$ yields
\begin{equation}
{ E_L^2 \over 2 \pi a} = J_L + 2n + \frac32 \ ,
\end{equation}
in agreement with Eq.~(\ref{eq:one}).

\section{Radial Dominance Analytic Solution}\label{sec:radial}

\subsection{Leading Order}

\begin{figure}[hbp]
\epsfxsize = \linewidth 
\hspace*{-5mm}
\epsfbox{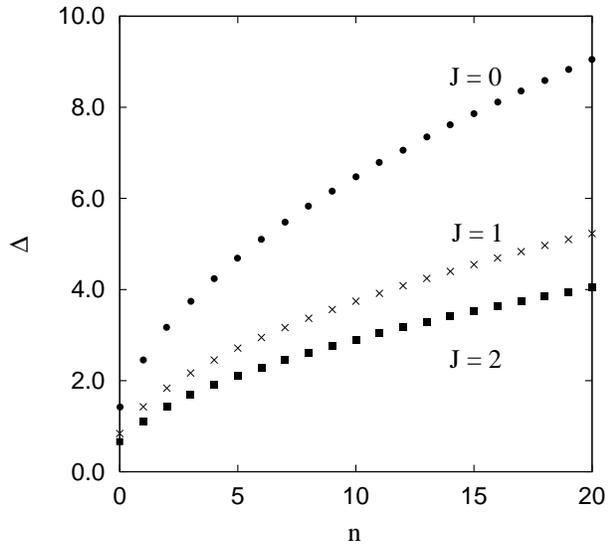}
\vskip 1 cm
\caption{The dependence of $\Delta$ upon the radial quantum number $n$ and
the angular momentum quantum number $J$. Increasing $n$ leads to
increasing $\Delta$.}
\label{fig:four}
\end{figure}

As verified by examining Fig.~\ref{fig:four}, large $n$ corresponds to
large $\Delta$ since at fixed angular momentum, both make the LDF energy
$E$ increase without bound.  By examining Eq.~(\ref{eq:wgam}), we see that
for large $\Delta$ either $\rho=-1$ or $v_\perp$ is small.  To verify this
we note that
\begin{equation}\label{eq:vSf}
v_\perp S(v_\perp) - f(v_\perp) = 2v_\perp/3 + O(v_\perp^3) \ .
\end{equation}
Hence, except at $\rho=-1$, $v_\perp \ll 1$ in the $\Delta \gg 1$ limit.

In the vanishing $v_\perp$ limit the expression (\ref{eq:w}) for $w(\rho)$
then becomes
\begin{equation}\label{eq:wLO}
w(\rho) \buildrel {\Delta \gg 1} \over \longrightarrow \Delta\Bigg[1 -
{2\over \pi}\big(1+\rho\big)\Bigg] \ .
\end{equation}
We note that the zero in $w$ that follows from Eq.~(\ref{eq:wLO}) is
\begin{equation}
\rho_+ = {\pi\over2} -1 \ .
\end{equation}

In Fig.~\ref{fig:five} we show the exact numerical result (solid curve) as well as the
large $\Delta$ approximation (LO) for $\Delta = 2$.

In this regime we may analytically evaluate the quantization integral
$I(\Delta)$ in Eq.~(\ref{eq:I}) to be 
\begin{equation}
I(\Delta) = \Delta \ .
\end{equation}
Substitution of this into Eq.~(\ref{eq:Qcond}) with the use of
Eq.~(\ref{eq:Deltatoo}) quickly yields
\begin{equation}\label{eq:spectrum}
{E^2 \over \pi a} = J + 2n + \frac32 \ .
\end{equation}
This is a good representation of the massless quark spectroscopy, as
we will discuss in the conclusion, Sec.~\ref{sec:conclusion}.
This is a reflection of the agreement of the
asymptotic (large $\Delta$) values of $I(\Delta)$ down to small $\Delta$.

Finally, we note that the spectrum in Eq.~(\ref{eq:spectrum}) is very
similar to the three dimensional harmonic oscillator spectrum as we have
pointed out previously in the context of the Lorentz scalar confinement of
massless quarks.

\subsection{Corrections to Leading Order}

\begin{figure}[hbp]
\epsfxsize = \linewidth 
\hspace*{-5mm}
\epsfbox{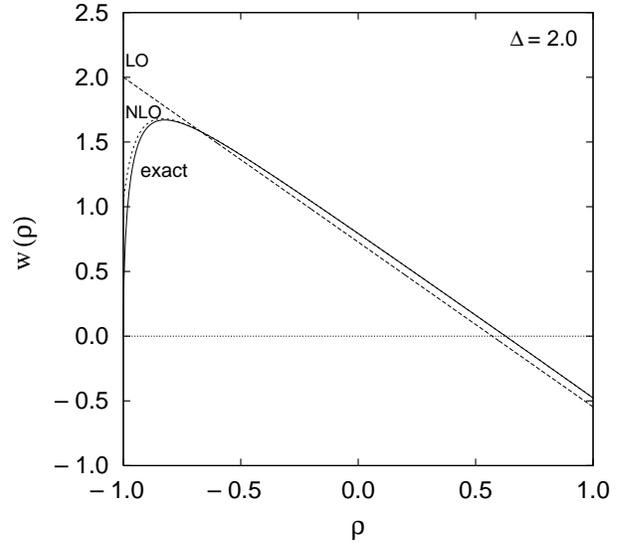}
\vskip 1 cm
\caption{A comparison of the exact numerical solution of
Eq.~(\protect\ref{eq:w}) for $w(\rho)$ (lowest curve) with the leading
order solution Eq.~(\protect\ref{eq:wLO}) (highest curve) and the
solution to next to leading order, Eq.~(\protect\ref{eq:wnlo2})
(intermediate curve). }
\label{fig:five}
\end{figure}

Although in leading order in $\Delta$ the result $I(\Delta) = \Delta$ is
very accurate, the distribution $w(\rho)$ computed to the same
approximation is not as satisfactory, which can be seen in
Fig.~\ref{fig:five}.  In this section we attempt to obtain a somewhat
better approximation to these two quantities.

Using Eq.~(\ref{eq:wgam}) and the limiting behavior of Eq.~(\ref{eq:vSf}),
we find the dependence of $v_\perp$ on $\rho$ and $\Delta$
\begin{eqnarray}\label{eq:vperpnlo}
v_\perp & = & \Bigg[2\left(\sqrt{1+\Delta^2} + \rho\Delta\right) \times \nonumber
\\ && \left(\sqrt{1+\Delta^2} - {4\over 3\pi} \left(\sqrt{1+\Delta^2} +
\rho\Delta\right)\right)\Bigg]^{-1} \ .
\end{eqnarray}

The value for upper turning point coordinate $\rho_+$ can be found from the
vanishing of Eq.~(\ref{eq:w}) for small $v_\perp$, which yields
\begin{equation}
\rho_+ = {\sqrt{1+\Delta^2}\over \Delta}\left({\pi\over2} - 1\right) \ .
\end{equation}
With this value of $\rho_+$, Eq.~(\ref{eq:vperpnlo}) determines the value
of $v_\perp$ at the upper turning point
\begin{equation}
v_\perp^+ = \left({3\over \pi}\right) {1\over 1 + \Delta^2} \ .
\end{equation}
To second order in $v_\perp$, $w(\rho)$ becomes
\begin{eqnarray}\label{eq:wnlo1}
w & \simeq & \sqrt{1+\Delta^2}\left(1 - {2\over \pi}\right) -
{2\rho\Delta\over\pi} \nonumber \\ &&
-\,{v_\perp^2\over2}\left[\sqrt{1+\Delta^2} - \,{4\over
3\pi}\left(\sqrt{1+\Delta^2} + \rho\Delta\right)\right] \ .
\end{eqnarray}
Upon substitution of the result of Eq.~(\ref{eq:vperpnlo}), $w(\rho)$
becomes
\begin{eqnarray}\label{eq:wnlo2}
w & \simeq & \sqrt{1+\Delta^2}\left(1 - {2\over \pi}\right) -
{2\rho\Delta\over\pi} \nonumber \\ &&
-\,{\left[\sqrt{1+\Delta^2} - \rho\Delta\right]^2\over 8
\left[\sqrt{1+\Delta^2} - \,{4\over 
3\pi}\left(\sqrt{1+\Delta^2} + \rho\Delta\right)\right]} \\
& \equiv & w_1 + w_2 \ , \nonumber
\end{eqnarray}
with $w_1$ being the piece of Eq.~(\ref{eq:wnlo1}) that is finite in the
limit of vanishing $v_\perp$.  The next-to-leading-order result is shown in
Fig.~\ref{fig:five}.  We see that it comes much closer to the exact
$w(\rho)$ but deviates slightly near $\rho = -1$ where the small $v_\perp$
expansion fails.

It is not difficult to evaluate the integral in Eq.~(\ref{eq:I}) by
changing the integration variable from $\rho$ to
\begin{eqnarray}
x & \equiv & \sqrt{1+\Delta^2} + \rho\Delta \ .
\end{eqnarray}
Splitting $I(\Delta) = I_1(\Delta) + I_2(\Delta)$ according to
Eq.~(\ref{eq:wnlo2}), we find
\begin{eqnarray}
I_1(\Delta) & = &
{1\over\Delta}\left({2\over\pi}\right)^2\left[\left({\pi\over 2} - 1\right)
\sqrt{1+\Delta^2} + \Delta \right]^2 , \\
I_2(\Delta) & = & {1\over 2\pi\Delta(1+\Delta^2)}\Bigg[{2\over\pi}- 1 - \Delta^2 -
\Delta\sqrt{1+\Delta^2} \nonumber \\  & - & 
{4\over3\pi}\ln\left({3\pi\Delta\over2}\left(\Delta +
\sqrt{1+\Delta^2}\right) + {3\pi -4 \over 2}\right)\Bigg] . \nonumber
\end{eqnarray}
Asymptotically $I(\Delta) = I_1(\Delta) + I_2(\Delta)$ is 
linear
\begin{equation}\label{eq:Deltaexp2}
I(\Delta)\simeq \Delta + \left(1 - {3\over
\pi}\right)\Delta^{-1} + O(\Delta^{-3} \ln(\Delta)) \ .
\end{equation}
Although the coefficient of $1/\Delta$ is small, yet higher order
corrections are expected to reduce it to the near zero value found in
(\ref{eq:numc}).

\section{Angular Dominance Analytic Solution} \label{sec:angular}

For large $E^2$ and small radial excitation, $\Delta$ is small as can be
seen from Eq.~(\ref{eq:Qcond}).  Since for a massless quark the leading
classical Regge trajectory corresponds to circular motion and in that case
$\Delta = 0$ and we are in the ultra-relativistic regime $v_\perp \approx
1$.  In this section we examine the behavior of Eqs.~(\ref{eq:wgam}) and
(\ref{eq:w}) as $v_\perp$ approaches unity.

The appropriate expansion parameter in this case is 
\begin{equation}
y = \sqrt{1 - v_\perp^2} = {1\over \gamma_\perp} \ ,
\end{equation}
which would be the radial velocity with a massless quark since a massless
quark must have $v^2 = \dot r^2 + v_\perp^2 = 1$.

\begin{figure}[hbp]
\epsfxsize = \linewidth 
\hspace*{-5mm}
\epsfbox{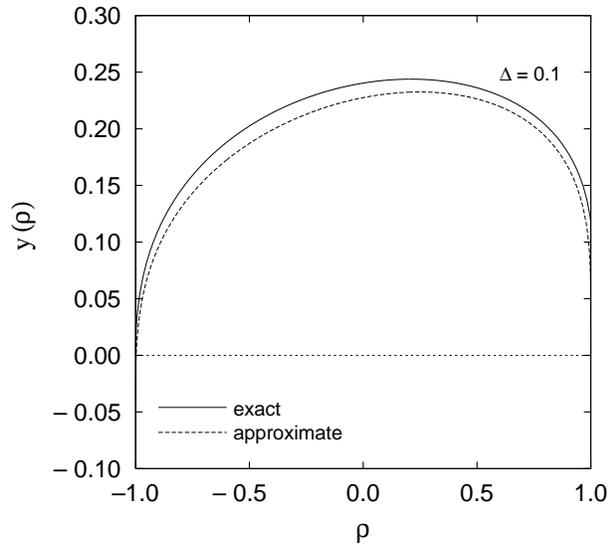}
\vskip 1 cm
\caption{A comparison of the exact numerical solution for $y(\rho) = (1-v_\perp^2)^{1/2}$ from
Eq.~(\protect\ref{eq:w}) and the solution of the cubic approximation,
Eq.~(\protect\ref{eq:cubic}), for $\Delta = 0.1$.}
\label{fig:six}
\end{figure}

Our first task is to expand Eq.~(\ref{eq:wgam}) for small $y$ and $\Delta$,
which is an expansion about circular orbits.  The result is
\begin{equation}\label{eq:cubic}
{8\over 3 \pi} y^3 - \rho\Delta y^2 - (1-\rho^2)\Delta^2 = 0 \ .
\end{equation}
The cubic term is critical.  Without it, there would be no real solutions
for $\rho^2 < 1$.  Going to quartic order in $y$ makes only minor changes.
To demonstrate the accuracy of this approximation, we show in
Fig.~\ref{fig:six} the solution $y(\rho)$ of Eq.~(\ref{eq:cubic}) and the
exact result obtained by solving the original (exact) Eq.~(\ref{eq:wgam})
for the value $\Delta = 0.1$.

For small $\Delta$ the $y(\rho)$ solution becomes more symmetric about
$\rho=0$.  The approximate solution to the cubic equation (\ref{eq:cubic})
for small $\Delta$ is
\begin{equation}\label{eq:yofrho}
y(\rho) \simeq \left({3\pi\over 8}\right)^{1/3} \left(1-\rho^2\right)^{1/3}
\Delta^{2/3} + {\pi\over 8}\rho\Delta + \ldots \ .
\end{equation}
By substitution of Eq.~(\ref{eq:yofrho}), we see that Eq.~(\ref{eq:cubic})
is satisfied up to the two leading powers of $\Delta$.  As $\Delta$ becomes
small the asymmetric term becomes of less relative importance and we may keep
only the leading term in Eq.~(\ref{eq:yofrho}),
\begin{equation}
y(\rho) \buildrel \Delta \ll 1  \over \longrightarrow \left({3\pi\over
8}\right)^{1/3} \left(1-\rho^2\right)^{1/3} \Delta^{2/3} \ .
\end{equation}

For $\Delta \ll 1$ and $y \ll 1$, the expression (\ref{eq:w})
for $w$ reduces to 
\begin{equation}\label{eq:wapprox}
w \simeq {2\over \pi} y^2 \simeq
{1\over2}\left({3\over\pi}\right)^{2/3}\left(1-\rho^2\right)^{2/3}
\Delta^{4/3} \ .
\end{equation}
With this approximation for $w$, the semi-classical
quantization integral (\ref{eq:I}) becomes
\begin{equation}
I = {4\over \pi}\int_{-1}^{+1} \kern -4pt d\rho\,\, w(\rho) = {8\over21}
\left({3\over \pi}\right)^{5/6} {\Gamma({2\over3})\over\Gamma({7\over6})}\,
\Delta^{4/3} \ .
\end{equation}
Numerically, we find
\begin{equation}\label{eq:Inumerical}
I(\Delta) \simeq 1.3367 \Delta^{4/3} \ .
\end{equation}

\begin{figure}[hbp]
\epsfxsize = \linewidth 
\hspace*{-5mm}
\epsfbox{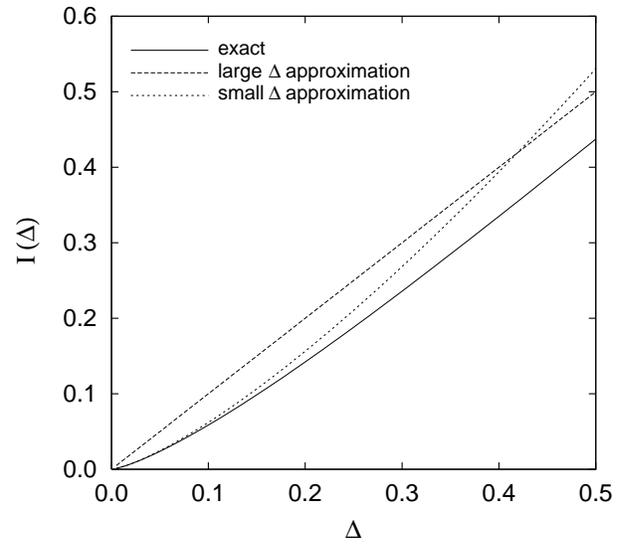}
\vskip 1 cm
\caption{A comparison of the exact numerical solution for $I(\Delta)$
(solid line) and the small $\Delta$ approximation (dotted line) and the
large $\Delta$ approximation (dashed line).}
\label{fig:seven}
\end{figure}

In Fig.~\ref{fig:seven} we show the small $\Delta$ values of $I(\Delta)$.
At small $\Delta$ we observe a distinct deviation from the linearity
observed at larger $\Delta$ (see Fig.~\ref{fig:two}).  The dotted curve
shown in Fig.~\ref{fig:seven} is the small $\Delta$ leading term
approximation of Eq.~(\ref{eq:Inumerical}) and the dashed curve is the
large $\Delta$ approximation to $I(\Delta)$.  Although the small $\Delta$
analytic approximation clearly is converging to the correct numerical
result for $\Delta \rightarrow 0$, deviations from the exact $I(\Delta)$
are evident around $\Delta=0.3$. For $\Delta > 0.4$, the large $\Delta$
approximation to $I(\Delta)$ is better.

\section{Conclusion} \label{sec:conclusion}

\begin{figure}[hbp]
\epsfxsize = \linewidth 
\hspace*{-5mm}
\epsfbox{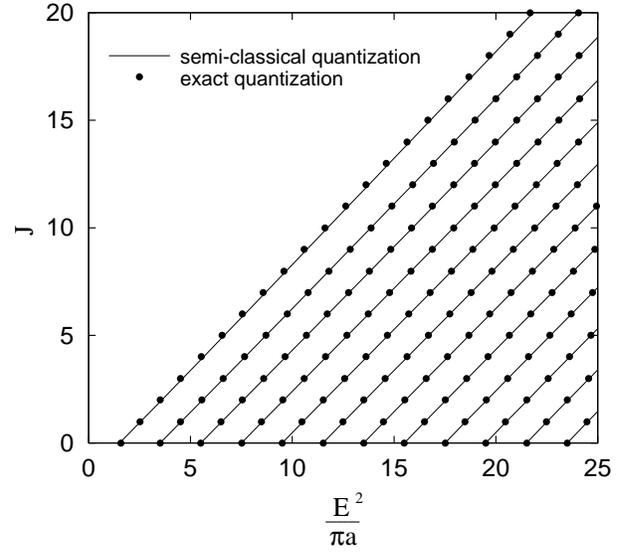}
\vskip 1 cm
\caption{The Regge trajectories from the semi-classical analysis (lines)
and by numerical canonical quantization (dots). }
\label{fig:three}
\end{figure}

\begin{figure}[hbp]
\epsfxsize = \linewidth 
\hspace*{-5mm}
\epsfbox{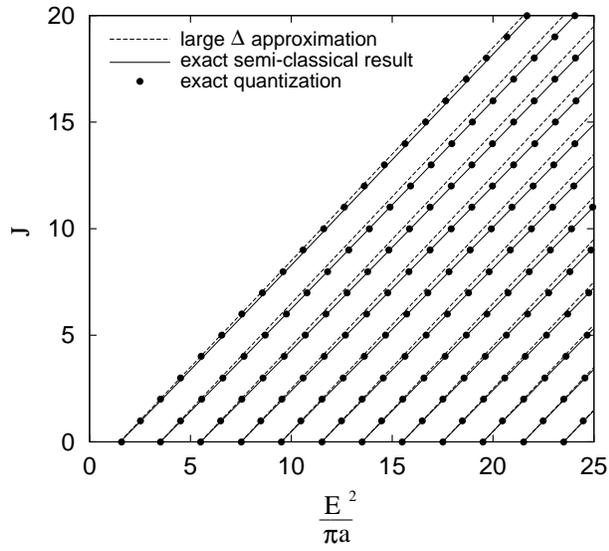}
\vskip 1 cm
\caption{The Regge trajectories from the large $\Delta$ approximation
(dashed lines) of Eq.~(1.1) in comparison to the exact numerical
semi-classical analysis (lines) and the results of numerical canonical
quantization (dots).}
\label{fig:eight}
\end{figure}

We have argued in Section \ref{sec:intro} that the straight string model of
confinement is accurate for ordinary hadron dynamics and that it is
therefore a serious candidate for long-range QCD.  Semi-classical
quantization of the straight string agrees well with a fully quantized
calculation \cite{ref:three}, as shown in Fig.~\ref{fig:three}.  The
semi-classical method allows considerable analytic insight into string
confinement in hadrons.  In particular, when one or two of the quarks are
massless, the entire spectroscopy is generated by a single integral
function $I(\Delta)$ given in Eq.~(\ref{eq:I}).  For almost all accessible
states $I(\Delta) \simeq \Delta$ which, by (\ref{eq:Qcond}), immediately
gives the simple pattern of Eq.~(\ref{eq:one}), as illustrated in
Fig.~\ref{fig:eight}.  Since even low-lying meson dynamics is dominated by
the confining region, the observation of degenerate towers of mesons of the
same parity becomes a key prediction of QCD.

The deviations from the simple pattern (\ref{eq:one}) are small and come
when $\Delta < 1$.  For observed meson states this corresponds to large
angular momentum and small radial excitation. That is, deviations from
(\ref{eq:one}) occur for mesons lying near the ``leading'' Regge
trajectory.  We showed in Sec.~\ref{sec:angular} that $I(\Delta)$ is
proportional to $\Delta^{4/3}$ for $\Delta \ll 1$ and the analytic
approximation agrees well with the exact numerical $I(\Delta)$, as can be
seen in Fig.~\ref{fig:seven}.

\begin{figure}[hbp]
\epsfxsize = \linewidth 
\hspace*{-5mm}
\epsfbox{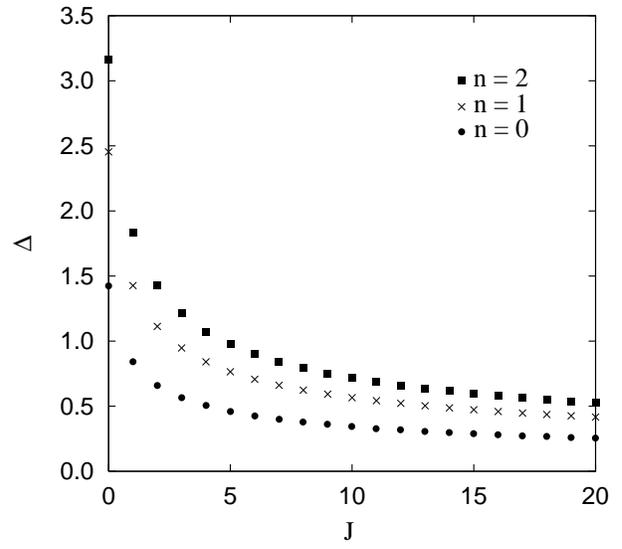}
\vskip 1 cm
\caption{The dependence of $\Delta$ upon the angular momentum quantum
number $J$ and the radial quantum number $n$. Increasing $J$ leads to
decreasing $\Delta$.}
\label{fig:nine}
\end{figure}

To make the above statement more explicit we show in Fig.~\ref{fig:eight} a
recreation of Fig.~\ref{fig:three} but now adding a dashed line
representing the analytic result (\ref{eq:spectrum}) which becomes exact
for large $n$ and small $J$.  For large $J$ and small $n$ the Regge
trajectories seem to have a slightly smaller slope.  In Fig.~\ref{fig:nine}
we show the $\Delta$ values in this orbital dominant regime and we see that
$\Delta$ is less than one.  As seen in Fig.~\ref{fig:seven}, $I(\Delta)<
\Delta$ and this reduction accounts for the difference between the dashed
and solid (exact) curves in Fig.~\ref{fig:eight}.

A final observation is that $I(\Delta) \rightarrow \Delta$ follows from the
radially dominant regime and not from the nearly circular approximation.
This limit automatically gives straight, evenly spaced Regge trajectories.

\section*{Acknowledgments}
This work was supported in part by the US Department of Energy under
Contract No.~DE-FG02-95ER40896.

\newpage

\newpage
\vspace{2cm}

\newpage
\vspace{2cm}

\newpage
\vspace{2cm}

\newpage
\vspace{2cm}

\newpage
\vspace{2cm}

\newpage
\vspace{2cm}

\newpage
\vspace{2cm}

\newpage
\vspace{2cm}

\newpage
\vspace{2cm}

\vfill

\end{document}